\documentclass[fleqn,11pt,twoside]{article}

\usepackage{amsthm,amsthm,amssymb, color, xcolor,epsfig, graphics, subfigure}

\usepackage{amsmath, graphicx, latexsym, lscape}

\newcommand{\nn}{\nonumber}

\newcommand{\be}{\begin{equation}}
\newcommand{\ee}{\end{equation}}
\newcommand{\bea}{\begin{eqnarray}}
\newcommand{\eea}{\end{eqnarray}}
\newcommand{\bse}{\begin{subequations}}
\newcommand{\ese}{\end{subequations}}

\DeclareMathAccent{\wtilde}{\mathord}{largesymbols}{"65}
\DeclareMathAccent{\what}{\mathord}{largesymbols}{"62}

\def\m@th{\mathsurround=0pt}
\mathchardef\bracell="0365
\def\upbrall{$\m@th\bracell$}
\def\undertilde#1{\mathop{\vtop{\ialign{##\crcr
    $\hfil\displaystyle{#1}\hfil$\crcr
     \noalign
     {\kern1.5pt\nointerlineskip}
     \upbrall\crcr\noalign{\kern1pt
   }}}}\limits}

\newcommand{\ub}[1]{\underline{#1}}
\newcommand{\wh}{\widehat}
\newcommand{\wt}{\widetilde}
\newcommand{\ut}{\undertilde}

\def\hypohat#1#2{\vrule depth #1 pt width 0pt{\smash{{\mathop{#2}
\limits_{\displaystyle\widehat{}}}}}}

\newcommand{\ol}{\overline}

\makeatletter
\newcommand{\copyrightnote}[2]{{\renewcommand{\thefootnote}{}
 \footnotetext{\small\it
\begin{flushleft}
 \copyright \ #1   #2
\end{flushleft}}}}

\newcommand{\Name}[1]{\begin{flushleft}
                       \LARGE \bf #1
                       \end{flushleft}\vspace{-3mm}}

\newcommand{\Author}[1]{\begin{flushleft}
                       \it #1 \end{flushleft}}

\newcommand{\Address}[1]{\begin{flushleft}
                       \it #1 \end{flushleft}}

\newcommand{\Date}[1]{\begin{flushleft}
                      \small  \it #1 \end{flushleft}}

\newcommand{\evenhead}{Author \ name}
\newcommand{\oddhead}{Article \ name}

\renewcommand{\@evenhead}{
\hspace*{-3pt}\raisebox{-15pt}[\headheight][0pt]{\vbox{\hbox to \textwidth
{\thepage \hfil \evenhead}\vskip4pt \hrule}}}
\renewcommand{\@oddhead}{
\hspace*{-3pt}\raisebox{-15pt}[\headheight][0pt]{\vbox{\hbox to \textwidth
{\oddhead \hfil \thepage}\vskip4pt\hrule}}}
\renewcommand{\@evenfoot}{}
\renewcommand{\@oddfoot}{}

\long\def\@makecaption#1#2{%
  \vskip\abovecaptionskip
  \sbox\@tempboxa{\small \textbf{#1.}\ \ #2}%
  \ifdim \wd\@tempboxa >\hsize
    {\small \textbf{#1.}\ \ #2}\par
  \else
    \global \@minipagefalse
    \hb@xt@\hsize{\hfil\box\@tempboxa\hfil}%
  \fi
  \vskip\belowcaptionskip}

\newcommand{\JNMPnumberwithin}[3][\arabic]{%
  \@ifundefined{c@#2}{\@nocounterr{#2}}{%
    \@ifundefined{c@#3}{\@nocnterr{#3}}{%
      \@addtoreset{#2}{#3}%
      \@xp\xdef\csname the#2\endcsname{%
        \@xp\@nx\csname the#3\endcsname .\@nx#1{#2}}}}%
}

\newcommand{\resetfootnoterule} {
  \renewcommand\footnoterule{%
  \kern-3\p@
  \hrule\@width.4\columnwidth
  \kern2.6\p@}
}

\renewcommand{\footnoterule}{}

\makeatother

\theoremstyle{definition}

\begin{document}

\renewcommand{\evenhead}{ {\LARGE\textcolor{blue!10!black!40!green}{{\sf \ \ \ ]ocnmp[}}}\strut\hfill F W Nijhoff and D-J Zhang}
\renewcommand{\oddhead}{ {\LARGE\textcolor{blue!10!black!40!green}{{\sf ]ocnmp[}}}\ \ \ \ \  Lagrangian multiform structure, extended lattice
Boussinesq system}

%%%% Matter for the first page
\thispagestyle{empty}
\newcommand{\FistPageHead}[3]{
\begin{flushleft}
\raisebox{8mm}[0pt][0pt]
{\footnotesize \sf
\parbox{150mm}{{Open Communications in Nonlinear Mathematical Physics}\ \ \ \ {\LARGE\textcolor{blue!10!black!40!green}{]ocnmp[}}
\quad Special Issue 1, 2024\ \  pp
#2\hfill {\sc #3}}}\vspace{-13mm}
\end{flushleft}}

\FistPageHead{1}{\pageref{firstpage}--\pageref{lastpage}}{ \ \ }

\strut\hfill

\strut\hfill

\copyrightnote{The author(s). Distributed under a Creative Commons Attribution 4.0 International License}

\begin{center}
%{\Large  {\sf This article is part of a Special Issue in Memory of Professor Decio Levi}}
{  {\bf This article is part of an OCNMP Special Issue\\
\smallskip
in Memory of Professor Decio Levi}}
\end{center}

\smallskip

\Name{On the Lagrangian multiform structure of the extended lattice Boussinesq system}

\Author{F.W. Nijhoff$^{\dagger\ddagger}$ and D-J. Zhang$^\ddagger$}

\Address{$\dagger$ School of Mathematics, University of Leeds, Leeds LS2 9JT, UK\\
$\ddagger$ Department of Mathematics, Shanghai University, Shanghai 200444, PR China}

\Date{Received December 30, 2023; Accepted January 26, 2024}

\setcounter{equation}{0}

\begin{abstract}

\noindent
The lattice Boussinesq (lBSQ) equation is a member of the lattice Gel'fand-Dikii (lGD) hierarchy, introduced in \cite{NijPapCapQui1992}, which is an infinite family of
integrable systems of partial difference equations labelled by an integer $N$, where
$N=2$ represents the lattice Korteweg-de Vries (KdV) system, and $N=3$ the Boussinesq system.
In \cite{Hiet2011} it was shown that, written as three-component system, the lBSQ system
allows for extra parameters which essentially amounts to building the lattice KdV
inside the lBSQ. In this paper we show that, on the level of the Lagrangian structure,
this boils down to a linear combination of Lagrangians from the members of the
lGD hierarchy as was established in \cite{LobbNijGD2010}. The corresponding Lagrangian multiform
structure is shown to exhibit a `double zero' structure.
\end{abstract}

\label{firstpage}
\hfill{\it Dedicated to the memory of Decio Levi}

%%%% The Article text starts here

\section{Introduction}

The extended lattice Boussinesq (lBSQ) system is given by the coupled system
of partial difference equations (P$\Delta$Es)
\begin{subequations}\label{eq:extBSQ}
\begin{align}
& \frac{\alpha_1(p-q)-\alpha_2(p^2-q^2)+\alpha_3(p^3-q^3)}{p-q+\wh{u}-\wt{u}} = \alpha_1-\alpha_2(p+q+u-\wh{\wt{u}}) \nn \\
& \qquad \qquad + \alpha_3\left[\wh{\wt{v}}-w +(p+q+u)(p+q-\wh{\wt{u}})-pq\right]\ ,
\label{eq:extBSQa} \\
& \wh{v}-\wt{v} = (p-q+\wh{u}-\wt{u})\wh{\wt{u}}+q\wt{u}-p\wh{u}\ , \label{eq:extBSQb} \\
& \wh{w}-\wt{w} =-(p-q+\wh{u}-\wt{u})u+p\wt{u}-q\wh{u}\ , \label{eq:extBSQc}
\end{align}
\end{subequations}
for the dependent variable fields $u=u(n,m)$, $v=v(n,m)$ and $w(n,m)$ depending on discrete
independent variables $n,m$ , where the $\wt{\phantom{a}}$ and $\wh{\phantom{a}}$ denote
elementary shifts on the lattice, i.e.
$$ \wt{u}=u(n+1,m), \quad \wh{u}=u(n,m+1), \quad \wh{\wt{u}}=u(n+1,m+1)\ , $$
and similarly for the fields $v$ and $w$. In \eqref{eq:extBSQ}, $p$ and $q$ are lattice parameters
associated with the variables $n$, $m$ respectively, whereas the $\alpha_i$, $i=1,2,3$
are fixed parameters. While the `pure' lattice BSQ equation (i.e., the case
that $\alpha_1=\alpha_2=0$)  was introduced in \cite{NijPapCapQui1992}, Hietarinta in
\cite{Hiet2011} by a systematic search of integrable cases found the extension of the lBSQ
system with the additional parameters. These extra parameters were subsequently
understood from the point of view of the `direct linearization method' in \cite{ZZN2012}
and extended to the entire lattice Gel'fand-Dikii (lGD) hierarchy. In fact, the
extra parameters arise from an unfolding of the underlying dispersion curve, which is
singular in the pure lBSQ case. Consequently, the soliton solutions associated with
the extended lBSQ system exhibit a physically more regular (forsooth non-singular) behaviour
as was demonstrated in e.g. \cite{HietZhang2011}.

In a separate development, the notion of Lagrangian multiforms was introduced in
\cite{LobNij2009} in order to provide a
variational framework for the phenomenon of multidimensional consistency (MDC),
(as exhibited for instance by the system \eqref{eq:extBSQ}): the key aspect of
integrability that multiple equations, in terms of a multitude of independent
variables, can be imposed on one and the same dependent variable such that
they allow for nontrivial common solutions (i.e. implying that those equations
are mutually compatible). In the Lagrangian multiform framework, the Lagrangians are
components of a differential- or difference $d$-form $\mathsf{L}$ ($d$ corresponding to the
dimensionality of the equations, i.e. the necessary minimal set of independent
variables appearing in each of the equations of the MDC system) and they are
integrated over arbitrary $d$-dimensional hypersurfaces in an embedding space of
arbitrary dimensionality to give the relevant action
functional $\mathcal{S}[\boldsymbol{u};\sigma]$ which is a functional of both the
field variables $\boldsymbol{u}=\boldsymbol{u}(\boldsymbol{x})$ (where $\boldsymbol{x}$ denotes
the set of independent variables) as well as of the hypersurfaces $\sigma$ in the space of
independent variables. The key new feature is that the least action principle is
to find the critical point for simultaneously varying the dependent veriables as well
as under deformations of the surfaces $\sigma$. This means that at the critical value
of the field variables the action is invariant under local deformations of the
surfaces of integration, which implies that they must obey a set of (extended) Euler-Lagrange (EL)
equations that possess the MDC property, as they must obey simultaneously compatible
EL equations on all choices of surfaces (subject to fixed boundary conditions).
Since its inception, this theory has been elaborated for many examples, in particular to the
case of the lGD hierarchy, \cite{LobbNijGD2010}, where a Lagrangian multiform structure
was presented for the entire hierarchy in terms of a single Lagrangian for each $N$.
In the present paper we focus on the case $N=3$ but show that the Lagrangians for the
hierarchy up to this level can be `summed up', in the sense of a linear combination of
Lagrangians, where the parameters $\alpha_i$ of the extended case emerge naturally.
We further demonstrate, going beyond the results of  \cite{LobbNijGD2010}, that this
Lagrangian multiform possesses a `double-zero structure' in the sense of
\cite{SleighNijCaud2020}, (cf. also \cite{SNC21,NijDarboux2023,MartinsNijRicc2023,CaudAttiSingh2023}) where it was shown that the exterior derivative of the
Lagrangian multiform for continuous MDC systems breaks down into sums of products of
factors which vanish on solutions of the EL equations\footnote{Similar
double-zero structures were recently exhibited by Vermeeren in the discrete case, cf.
\cite{Verm2023}.}.

\section{The extended lattice Boussinesq Lagrangians}

A Lagrangian for the pure lattice Boussinesq equation in scalar form, which is a 9-point
equation for the field $u$, obtained by eliminating $v$ and $w$ from the system \eqref{eq:extBSQ},
was given in \cite{NijPapCapQui1992}.
Associated 9-point lattice systems are the lattice modified Boussinesq (lMBSQ) equation,
derived in the same paper, and lattice Schwarzian Boussinesq (lSBSQ) equation, \cite{NijSchwarz1996},
while the parameter extension (coined NQC (Nijhoff-Quispel-Capel) type BSQ equation, as it
can be thought of as the rank 3 version of the primary
lattice equation found in \cite{NQC1983}) was found in \cite{ZZN2012}, cf. also
\cite{NijSunZhang2022}. From these higher-order equations which go beyond the well-studied
case of quad equations, only for the lMBSQ a Lagrangian structure has so far been
found, \cite{AtkLobbNij2012}, while for the Schwarzian variants of the lBSQ no Lagrangian structure
so far exists.
\iffalse
\footnote{In the continuous case the associated continuous `generating PDE' for the
SBSQ is similar to the Lagrangian for the corresponding regular BSQ case, \cite{TonNij2005ii},
while the Lagrangian structure for the generating PDE of the MBSQ system was also found
\cite{LobbNijtbp}.}.
\fi

The Lagrangian for the lBSQ we will consider here, reads
\bea\label{eq:LBSQLagr}
&& \mathcal{L}^{(3)}_{pq} = (p^3-q^3)\ln(p-q+\wh{u}-\wt{u}) +(p^2+pq+q^2)(\wh{u} -\wt{u}) \nn \\
&& - (p+q+u)(p+q-\wh{\wt{u}})(p-q+\wh{u}-\wt{u}) +\wh{\wt{u}}(q\wt{u}-p\wh{u})\   ,
\eea
which differs from the Lagrangian given in \cite{NijPapCapQui1992} through the presence of
some linear difference terms, which are unimportant for the conventional EL equations, leading
to a the 9-point scalar equation which is the lBSQ.
However, these total difference terms are important for the so-called \textit{corner equations}
and for the `double-zero' structure of the lBSQ system (in analogy to the double-zero
phenomenon that appeared in the continuous Lagrangian multiform structures, cf. \cite{SleighNijCaud2020}).
We supplement the lBSQ Lagrangian \eqref{eq:LBSQLagr} with the Lagrangian for the lattice
(potential) Korteweg-de Vries (lKdV) in the following form
\be\label{eq:LKdVLagr}
\mathcal{L}^{(2)}_{pq} = (q^2-p^2)\ln(p-q+\wh{u}-\wt{u}) +
u(\wh{u} -\wt{u})+(p-q)(u-\wh{\wt{u}})\ .
\ee
Here also we include linear terms, which do not contribute to the usual discrete EL equations,
in comparison to the original lKdV Lagrangian that was first found in \cite{CapNijPap1991},
again for reasons mentioned above. Furthermore, we also add the `trivial' Lagrangian
\be\label{eq:TrivLagr}
\mathcal{L}^{(1)}_{pq} = (p-q)\ln(p-q+\wh{u}-\wt{u})\ ,
\ee
the EL equation of which leads to the linear P$\Delta$E:  $\wh{\ut{u}}+\wt{\hypohat 0 u}-2u=0$,
with the under-accents denoting the backward shifts.

It turns out that the linear combination of these Lagrangians
\be \label{eq:extBSQLagr}
\mathcal{L}_{pq}=\alpha_1 \mathcal{L}^{(1)}_{pq}+\alpha_2\mathcal{L}^{(2)}_{pq}+\alpha_3\mathcal{L}^{(3)}_{pq}
\ee
forms the Lagrangian for the extended lBSQ system, whose conventional EL equation leads to the 9-point
equation that is obtained from the system \eqref{eq:extBSQ} by eliminating the fields $v$ and $w$:
\begin{eqnarray}\label{eq:fullBSQ}
&& \frac{P-Q}{p-q+u-\wt{\hypohat 0 u}}-\frac{P-Q}{p-q+\wh{\ut{u}}-u} =
(\alpha_3(p+q)-\alpha_2)\left(\wh{u}+{\hypohat 0 u}-\wt{u}-\ut{u}\right) \nn \\
&& +\alpha_3\left[ \ut{u}\wh{u}-{\hypohat 0 u}\wt{u}+p( \wh{u}+{\hypohat 0 u})
-q(\wt{u}+\ut{u})  \right. \nn \\
&& \left. - (p-q+\wh{u}-\wt{u})\wh{\wt{u}}- (p-q+\ut{u}-{\hypohat 0 u})\ut{{\hypohat 3 u}}\right]\ ,
\end{eqnarray}
in which $P=\alpha_1 p-\alpha_2 p^2+\alpha_3 p^3$, $Q=\alpha_1 q-\alpha_2 q^2+\alpha_3 q^3$. So far this establishes
the conventional Lagrangian structure for the extended lBSQ system. We now proceed to
establishing the multiform structure.

\section{The extended lattice Boussinesq Lagrangian multiform structure}

For 2-dimensional integrable lattice equations we expect a discrete Lagrangian 2-form
structure which can be written as
\be\label{eq:LBSQmulti}
\mathsf{L}= \sum_{i<j} \mathcal{L}_{p_ip_j} \delta_{p_i}\wedge\delta_{p_j}\  ,
\ee
where the $\mathcal{L}_{p_ip_j}$ are the Lagrangian components for any two directions indicated by
the lattice parmeters $p_i$, $p_j$ of multidimensional regular lattice of, in principle, arbitrary
dimension.  The parameters $p$ and $q$ of the previous section are just two possible
choices for the parameters $p_i$, among many additional parameters, and with each parameter
there is a discrete lattice variable $n_{p_i}$  playing the role of coordinates for the $i^{\rm th}$
direction in that multidimensional lattice. In \eqref{eq:LBSQmulti}
the $\delta_p$ denotes a \textit{discrete differential}\footnote{The notation is similar to the
one introduced in \cite{MansHyd2012}.}, i.e. a formal symbol indicating that in the action
functional
\be\label{eq:action}
S[u(\boldsymbol{n});\sigma]=\sum_\sigma \mathsf{L}= \sum_{\sigma_{ij}\in\sigma}
\mathcal{L}_{p_ip_j} \delta_{p_i}\wedge\delta_{p_j} ,
\ee
the Lagrangian contributions from all elementary quads $\sigma_{ij}=(\boldsymbol{n},\boldsymbol{n}+\boldsymbol{e}_i,
\boldsymbol{n}+\boldsymbol{e}_j)$ (with elementary displacement vectors $\boldsymbol{e}_i$
along the edges  in the lattice associated with the lattice parameter $p_i$) are simply
summed up according to their base point $\boldsymbol{n}$ and their orientation.

According to the derivation proposed in \cite{LobNij2018}, the set of multiform EL equations is
obtained by considering the smallest closed quad-surface, which is simply an elementary cube,
and the action of which is given by
\be\label{eq:boxL}
S[u(\boldsymbol{n});{\rm cube}]=:(\square\mathcal{L})_{pqr}=
\Delta_p\mathcal{L}_{qr}+ \Delta_q\mathcal{L}_{rp}+\Delta_r\mathcal{L}_{pq}\ ,
\ee
where $\Delta_p=T_p-{\rm id}$ is the forward difference operator in the direction labeled by $p$, and
where $T_p$ denotes the elementary forward shift operator in that direction, i.e.
$$ T_pu(\cdots,n_p,\cdots)=u(\cdots,n_p+1,\cdots)=u(\boldsymbol{n}+\boldsymbol{e}_p)\  . $$
The fundamental EL equations of the multiform are obtained by taking partial derivatives
w.r.t. all the internal vertices, i.e. the variables $u$, $\wt{u}$, $\wh{u}$, $\ol{u}$,
which represent the shifts in all three directions $n_p, n_q, n_r$, and w.r.t. the variable with the combined
shifts $\wh{\wt{u}}$, $\wh{\ol{u}}$, $\wt{\ol{u}}$, as well as w.r.t. the triply shifted $\wh{\wt{\ol{u}}}$.
With the cube action for the lBSQ ($N=3$) contribution to the multiform calculated as
\begin{align*}
&(\square\mathcal{L}^{(3)})_{pqr} =
(p^3-q^3)\ln\left(\frac{p-q+\wh{\ol{u}}-\wt{\ol{u}}}{p-q+\wh{u}-\wt{u}}\right)
%+\wh{\wt{\ol{u}}}\left(q\wt{\ol{u}}-p\wh{\ol{u}}\right)
-\wh{\wt{u}}(q\wt{u}-p\wh{u}) + {\rm cycl.} \\
& -(p^2+pq+q^2)(\wh{u}-\wt{u}) - (p-q+\wh{\ol{u}}-\wt{\ol{u}})(p+q+\ol{u})(p+q-\wh{\wt{\ol{u}}})+ {\rm cycl.}  \\
& + (p^2+pq+q^2)(\wh{\ol{u}}-\wt{\ol{u}})+(p-q+\wh{u}-\wt{u})(p+q+u)(p+q-\wh{\wt{u}})+{\rm cycl.}\ ,
\end{align*}
where $+ {\rm cycl.}$ means adding two similar terms after cyclic permutations of $p,q,r$ and
the respective shifts $\wt{\phantom{a}},\wh{\phantom{a}},\ol{\phantom{a}}$.
The contribution to the action from the lKdV components reads:
\begin{align*}
(\square\mathcal{L}^{(2)})_{pqr} =
(q^2-p^2)\ln\left(\frac{p-q+\wh{\ol{u}}-\wt{\ol{u}}}{p-q+\wh{u}-\wt{u}}\right)
+ \ol{u}(\wh{\ol{u}}-\wt{\ol{u}}) +(p-q)( \ol{u}+\wh{\wt{u}}) + {\rm cycl.} .
\end{align*}
Varying independently w.r.t. the variables $u$ at all the internal vertices of the closed surface
of the cube leads to the corner equations. Let us, for simplicity, treat the corner equations for
the lKdV and lBSQ separately. First, for the lKdV component
(setting here $\alpha_1=\alpha_3=0$, $\alpha_2=1$) we have as only nontrivial contributions
\begin{subequations}
\begin{align}
\frac{\partial(\square\mathcal{L}^{(2)})_{pqr}}{\partial \ol{u}} &=
\left( \wh{\ol{u}}-q+\frac{q^2-r^2}{q-r+\ol{u}-\wh{u}}\right)
-\left( \wt{\ol{u}}-p+\frac{p^2-r^2}{p-r+\ol{u}-\wt{u}}\right)=0\ ,\label{eq:CELe}\\
\frac{\partial(\square\mathcal{L}^{(2)})_{pqr}}{\partial \wh{\wt{u}}} &=
\left( -\wt{u}-q+\frac{q^2-r^2}{q-r+\wt{\ol{u}}-\wh{\wt{u}}}\right)
-\left( -\wh{u}-p+\frac{p^2-r^2}{p-r+\wh{\ol{u}}-\wh{\wt{u}}}\right)=0\ .
\label{eq:CELf}
\end{align}
\end{subequations}
Since \eqref{eq:CELe} and \eqref{eq:CELf} must hold for every $p$, $q$ (and corresponding
lattice shifts) while fixing $r$, it is evident that we deduce the two conditions
$$ \wh{\ol{u}}-q+\frac{q^2-r^2}{q-r+\ol{u}-\wh{u}}=f_r, \quad -\wt{u}-q+\frac{q^2-r^2}{q-r+\wt{\ol{u}}-\wh{\wt{u}}}=\wh{\wt{g}}_r,   $$
where $f_r$ and $g_r$ are independent of $p$, $q$ and their corresponding lattice shifts,
the only consistent choice being $f_r=r_0+u$, $g_r= r_0-\ol{u}$ (where $r_0$ is an arbitrary
constant which may only depend on the parameter $r$), and
which leads to the lattice potential KdV equation as a quad-lattice equation.

Let us next consider the corner equations for the lBSQ components $\mathcal{L}^{(3)}$ of the multiform. In what follows we will
use the abbreviations
$$
\Gamma_{pq}:=p-q+T_qu-T_pu,  \quad
\Gamma_{qr}:=q-r+T_ru-T_qu,  \quad
\Gamma_{rp}:=r-p+T_pu-T_ru,
$$
with $T_pu=\wt{u}$, $T_qu=\wh{u}$, $T_ru=\ol{u}$
and denote
\begin{align*}
\Gamma_{pqr}&= \Gamma_{pq}(r+T_pT_qu)+  \Gamma_{qr}(p+T_qT_ru)+\Gamma_{rp}(q+T_rT_pu) \\
 &= (T_r\Gamma_{pq})(T_ru-r)+  (T_p\Gamma_{qr})(T_pu-p)+(T_q\Gamma_{rp})(T_qu-q)\  ,
\end{align*}
where we note that $\Gamma_{pqr}=0$ is actually the lattice Kadomtsev-Petviashvili (lKP) equation
\cite{NijCapWieQui1984}, which holds for the lBSQ solutions as a consequence of
\eqref{eq:extBSQb} and  \eqref{eq:extBSQc}.
From the elementary cube action
$(\square\mathcal{L}^{(3)})_{pqr}$ we obtain the following corner relations:
\begin{subequations} \label{eq:CEL}
\begin{align}
\frac{\partial(\square\mathcal{L}^{(3)})_{pqr}}{\partial u} &= -\Gamma_{pqr}=0 , \label{eq:CELa} \\
\frac{\partial(\square\mathcal{L}^{(3)})_{pqr}}{\partial \wh{\wt{\ol{u}}}} &= \Gamma_{pqr}=0 , \label{eq:CELb} \\
\frac{\partial(\square\mathcal{L}^{(3)})_{pqr}}{\partial \ol{u}} &= \frac{r^3-p^3}{r-p+\wt{u}-\ol{u}}
-\frac{q^3-r^3}{q-r+\ol{u}-\wh{u}} -p\wt{\ol{u}}+q\wh{\ol{u}} \nn \\
& \quad  +(p-q+\wh{\ol{u}}-\wt{\ol{u}})(p+q-\wh{\wt{\ol{u}}})
+(r^2+rp+p^2)-(q^2+rq+r^2) \nn \\
& \quad +(q+r+u)(q+r-\wh{\ol{u}})
-(r+p+u)(r+p-\wt{\ol{u}})=0\ , \label{eq:CELc}\\
\frac{\partial(\square\mathcal{L}^{(3)})_{pqr}}{\partial \wh{\wt{u}}} &=
\frac{r^3-p^3}{r-p+\wh{\wt{u}}-\wh{\ol{u}}}
-\frac{q^3-r^3}{q-r+\wt{\ol{u}}-\wh{\wt{u}}} +p\wh{u}-q\wt{u} \nn \\
& \quad  -(p-q+\wh{u}-\wt{u})(p+q+u)
+(r^2+rp+p^2)-(q^2+rq+r^2) \nn \\
& \quad +(q+r+\wt{u})(q+r-\wh{\wt{\ol{u}}})
-(r+p+\wh{u})(r+p-\wh{\wt{\ol{u}}})=0\ , \label{eq:CELd}
\end{align}
\end{subequations}
(and similar equations to \eqref{eq:CELc} and \eqref{eq:CELd} upon cyclic permutations) which form essentially the system of EL equations for the multiform action. We note that using the lKP equation
$\Gamma_{pqr}=0$ we can rewrite \eqref{eq:CELc} as
\begin{align*}
& \left[r^2+rp+p^2-p\wt{\ol{u}} +\frac{p^3-r^3}{p-r+\ol{u}-\wt{u}} -(r+p+u)(r+p-\wt{\ol{u}})
-(r+p-\wt{\ol{\ol{u}}})(p-r+\ol{\ol{u}}-\wt{\ol{u}})\right] \\
& -\left[r^2+rq+q^2-q\wh{\ol{u}} +\frac{q^3-r^3}{q-r+\ol{u}-\wh{u}} -(r+q+u)(r+q-\wh{\ol{u}})
-(r+q-\wh{\ol{\ol{u}}})(q-r+\ol{\ol{u}}-\wh{\ol{u}})\right]=0
\end{align*}
and \eqref{eq:CELd} as
\begin{align*}
& \left[r^2+rp+p^2+p\wh{u} +\frac{p^3-r^3}{p-r+\wh{\ol{u}}-\wh{\wt{u}}} -(r+p+\wh{u})(r+p-\wh{\wt{\ol{u}}})
+(r+p+\wh{\ub{u}})(r-p+\wh{\wt{\ub{u}}}-\wh{u})\right] \\
& -\left[r^2+rq+q^2+q\wt{u} +\frac{q^3-r^3}{q-r+\wt{\ol{u}}-\wh{\wt{u}}} -(r+q+\wt{u})(r+q-\wh{\wt{\ol{u}}})
+(r+q+\wt{\ub{u}})(r-q+\wh{\wt{\ub{u}}}-\wt{u})\right]=0\ ,
\end{align*}
which hold for all $p$, $q$ and their corresponding shifts, while fixing $r$.
Thus we conclude that the following relations hold:
\begin{align*}
& r^2+rq+q^2-q\wh{\ol{u}} +\frac{q^3-r^3}{q-r+\ol{u}-\wh{u}} -(r+q+u)(r+q-\wh{\ol{u}})
-(r+q-\wh{\ol{\ol{u}}})(q-r+\ol{\ol{u}}-\wh{\ol{u}})=F_r \\
& r^2+rq+q^2+q\wt{u} +\frac{q^3-r^3}{q-r+\wt{\ol{u}}-\wh{\wt{u}}} -(r+q+\wt{u})(r+q-\wh{\wt{\ol{u}}})
+(r+q+\wt{\ub{u}})(r-q+\wh{\wt{\ub{u}}}-\wt{u})=\wt{\wh{G}}_r
\end{align*}
for all lattice parameters $q$ and corresponding lattice shifts $\wh{\phantom{a}}$~, where $F_r$ and $G_r$ are independent
of $q$ and the associated shifts and only depends on $r$ and may only involve (single or multiple) ~$\ol{\phantom{a}}$~ shifts
acting on $u$. In fact, using \eqref{eq:extBSQ} we have the identifications
$$ F_r=2r^2+\ol{\ol{v}}-w-r\ol{\ol{u}}\ , \quad G_r=2r^2+\ol{v}-\ub{w}+r\ub{u}\  , $$
but since the quantities $v$ and $w$ are not present in the Lagrangians, in these formulas the quantity $\ol{\ol{v}}-w$ can be
considered as a potential field.  Eliminating this potential field, the ensuing relation ~$G_r-r\ub{u}=\ub{F}_r+r\ol{u}$~ leads to the
9-point lBSQ equation \eqref{eq:fullBSQ} (for the choice made here, $\alpha_1=\alpha_2=0$, $\alpha_3=1$).

We will now elucidate the double-zero
structure of the lBSQ multiform.\footnote{Similar double-zero structures have been
recently established in \cite{RichVerm2023} for the Lagrangian multiforms for the well-known Adler-Bobenko-Suris (ABS, cf. \cite{AdlBobSur2002})  list of integrable
quad-equations. Lagrangians for those equations are based on 3-leg formulae
for those quad equations, but these may not be universal for higher rank systems like the lBSQ system, and
the latter does not share the same symmetries of the square.
We note that the specific $log$ structure in the Lagrangians, exploited here, holds for Lagrangians of the entire lGD hierarchy,
and indeed also for the Lagrangians 3-forms for the lattice and semi-discrete
versions of the KP equation, cf. \cite{NijKP2023}.}
The double-zero structure for the extended lBSQ multiform is based on the following identities:
\begin{subequations} \label{eq:DoubleZero}
\begin{align}\label{eq:DoubleZeroBSQ}
(\square\mathcal{L}^{(3)})_{pqr} =& -(p+q+r+u-\wh{\wt{\ol{u}}})\Gamma_{pqr} \nn \\
&+ p^3\ln\left(\frac{\ol{\Gamma}_{pq}}{\Gamma_{pq}}\cdot
\frac{\Gamma_{rp}}{\wh{\Gamma}_{rp}}\right)
+ q^3\ln\left(\frac{\wt{\Gamma}_{qr}}{\Gamma_{qr}}\cdot
\frac{\Gamma_{pq}}{\ol{\Gamma}_{pq}}\right)
+ r^3\ln\left(\frac{\wh{\Gamma}_{rp}}{\Gamma_{rp}}\cdot
\frac{\Gamma_{qr}}{\wt{\Gamma}_{qr}}\right) 
\end{align}
for the lBSQ components and
\begin{align}\label{eq:DoubleZeroKdV}
(\square\mathcal{L}^{(2)})_{pqr} =& \Gamma_{pqr}
- p^2\ln\left(\frac{\ol{\Gamma}_{pq}}{\Gamma_{pq}}\cdot
\frac{\Gamma_{rp}}{\wh{\Gamma}_{rp}}\right)
- q^2\ln\left(\frac{\wt{\Gamma}_{qr}}{\Gamma_{qr}}\cdot
\frac{\Gamma_{pq}}{\ol{\Gamma}_{pq}}\right)
- r^2\ln\left(\frac{\wh{\Gamma}_{rp}}{\Gamma_{rp}}\cdot
\frac{\Gamma_{qr}}{\wt{\Gamma}_{qr}}\right) 
\end{align}
for the lKdV components, 
while for the linear components we have
\begin{align}\label{eq:DoubleZeroLin}
(\square\mathcal{L}^{(1)})_{pqr} =
p\ln\left(\frac{\ol{\Gamma}_{pq}}{\Gamma_{pq}}\cdot
\frac{\Gamma_{rp}}{\wh{\Gamma}_{rp}}\right)
+ q\ln\left(\frac{\wt{\Gamma}_{qr}}{\Gamma_{qr}}\cdot
\frac{\Gamma_{pq}}{\ol{\Gamma}_{pq}}\right)
+ r\ln\left(\frac{\wh{\Gamma}_{rp}}{\Gamma_{rp}}\cdot
\frac{\Gamma_{qr}}{\wt{\Gamma}_{qr}}\right)\  .
\end{align}
\end{subequations}
In eqs. \eqref{eq:DoubleZero} the shifted functions $\Gamma_{pq}$
are expressed as
$$
\ol{\Gamma}_{pq}=p-q+\wh{\ol{u}}-\wt{\ol{u}}\ , \quad
\wt{\Gamma}_{qr}=q-r+\wt{\ol{u}}-\wh{\wt{u}}\ , \quad
\wh{\Gamma}_{rp}=r-p+\wh{\wt{u}}-\wh{\ol{u}}\ ,
$$
where $\wh{\ol{u}}=T_qT_ru$, $\wh{\wt{u}}=T_qT_pu$, $\wt{\ol{u}}=T_pT_ru$.

Now it is noted that both $\Gamma_{pqr}$ and the factors within the logarithms,
have a zero, resp. a logarithmic zero (i.e. where the factors equal unity) on solutions of
the lKP equation. In fact, due to the identities:
$$ \Gamma_{pqr}=\ol{\Gamma}_{pq}\Gamma_{qr}-\Gamma_{pq}\wt{\Gamma}_{qr}
= \wt{\Gamma}_{qr}\Gamma_{rp}-\Gamma_{qr}\wh{\Gamma}_{rp}
= \wh{\Gamma}_{rp}\Gamma_{pq}-\Gamma_{rp}\ol{\Gamma}_{pq}\ , $$
we can factorise the exterior derivative of  the extended Lagrangian multiform
components \eqref{eq:extBSQLagr} as follows
$$ (\square\mathcal{L})_{pqr}=\Gamma_{pqr}\times \left[ \alpha_2-\alpha_3(p+q+r+u-\wh{\wt{\ol{u}}})+
\frac{1}{\Gamma_{pqr}}\left(P\ln\left(1-\frac{\Gamma_{pqr}}{\Gamma_{pq}
\wh{\Gamma}_{rp}} \right) +{\rm cycl.}  \right)   \right] \ ,
$$
where the terms from the logarithms in the second factor have a zero for $\Gamma_{pqr}=0$.
Thus, the multiform variational equations from $\delta(\square\mathcal{L})_{pqr}=0$,
yield two equations
\begin{subequations}\label{eq:DZ}\begin{align}
& {\rm i)}  ~\quad \Gamma_{pqr}=0\  , \label{eq:DZi} \\
& {\rm ii)} \quad \alpha_2+\alpha_3\wh{\wt{\ol{u}}}= \alpha_3(p+q+r+u) +
\frac{\Gamma_{rp}}{\wh{\Gamma}_{rp}}\left(
\frac{P}{\Gamma_{pq}\Gamma_{rp}}+\frac{Q}{\Gamma_{qr}\Gamma_{pq}}
+\frac{R}{\Gamma_{rp}\Gamma_{qr}}\right)\  , \label{eq:DZii}
\end{align} \end{subequations}
(in which in addition to $P$, $Q$, introduced earlier, we have
$R=\alpha_1 r-\alpha_2r^2+\alpha_3 r^3$)
where we have used in the expansions of the logarithms in the factor above that the higher
order terms do not contribute on solutions of the lKP equation, $\Gamma_{pqr}=0$, and
that the prefactor in the second term on the r.h.s. of \eqref{eq:DZii} is invariant
under cyclic permutations of the indices (again on solutions of the lKP equation).
We can perhaps consider the second equation, \eqref{eq:DZii}, as in some sense a BSQ `dual' to the
first equation.

From eqs. \eqref{eq:DZ} a non-potential extended lBSQ system
can be deduced in the following way.
Fix $r$ and the shift $\ol{u}$, and introduce  the quantities $\Gamma_p:=\Gamma_{pr}$
and $\Gamma_q:=\Gamma_{qr}$. Then in terms of $\Gamma_p,\Gamma_q$ we have the following
coupled two-dimensional lattice system:
\begin{subequations}\begin{align}
& \frac{\wh{\Gamma}_p}{\Gamma_p}=\frac{\wt{\Gamma}_q}{\Gamma_q} \ ,\\
& \alpha_3\left(\wh{\wh{\wt{\Gamma}}}_p-\wh{\wt{\wt{\Gamma}}}_q- \Gamma_p+\Gamma_q\right)
=\frac{\wh{\Gamma}_q}{\wh{\wt{\Gamma}}_q}\left(\frac{P}{\wh{\Gamma}_p(\wh{\Gamma}_q-\wh{\Gamma}_p)}
+\frac{Q}{\wh{\Gamma}_q(\wh{\Gamma}_p-\wh{\Gamma}_q)}-\frac{R}{\wh{\Gamma}_p\wh{\Gamma}_q} \right) \nn \\
& \qquad\qquad  -  \frac{\wt{\Gamma}_p}{\wh{\wt{\Gamma}}_p}\left(\frac{P}{\wt{\Gamma}_p(\wt{\Gamma}_q-\wt{\Gamma}_p)}
+\frac{Q}{\wt{\Gamma}_q(\wt{\Gamma}_p-\wt{\Gamma}_q)}-\frac{R}{\wt{\Gamma}_p\wt{\Gamma}_q} \right)\ ,
\end{align}\end{subequations}
which may play the same role as the non-potential KdV equation, cf. \cite{HJN16}, relative to the potential KdV quad-equation.

 \section{Discussion}

It is almost evident that the Lagrangian multiform structure for the whole extended
lGD hierarchy can be obtained by taking the earlier results from \cite{LobbNijGD2010}, and
use the collection of Lagrangians for any $N$ to construct the linear combination
$$ \mathcal{L}_{pq} =\sum_{N=1}^M \alpha_N\mathcal{L}^{(N)}_{pq}\ ,  $$
and do a resummation of the $p$,$q$-dependent factors therein. The resulting
Lagrangian structure will necessarily have closure, as a consequence of the closure
relation that was proven in the 2010 paper \cite{LobbNijGD2010}. A technical difficulty is that
as we climb in the lGD hierarchy the number of component fields will increase, while the
Lagrangian components $\mathcal{L}^{(N)}$ for $N=1,2,3$ as presented here are
in terms of one scalar field. Furthermore, in \cite{LobbNijGD2010} the notion of
corner equations was not yet established, so some additional linear terms in the Lagrangians
will need to be computed. We will postpone that work to a future publication.
Furthermore, we speculate that in the limit that $M\to\infty$, i.e. the infinite-component
case of the lGD hierarchy, we may expect essentially that a Lagrangian multiform
structure for the lKP system itself will appear. Even though, the equation $\Gamma_{pqr}=0$
that appears at all levels in the lGD Lagrange structure, which is already the
lKP equation, one should note that in this context it is constrained by the additional equations,
which renders the solutions as essentially that of a 2-dimensional lattice field theory.
However, in the limit $M\to\infty$ we expect to regain the unconstrained lKP system,
which can be considered as a true 3-dimensional lattice field theory\footnote{An alternative
formulation of the Lagrangian structure for the lKP system is developed in \cite{NijKP2023}.}.

\subsection*{Acknowledgements}

The research was supported by NSFC grants (Nos. 12171306, 12271334)
and National Foreign Expert Program of China (No. G2022172028L).
% \todo{acknowledgement of relevant grants!}\\
FWN is indebted to discussions with J. Richardson and M. Vermeeren regarding the
double-zero structure in the discrete case.

\label{lastpage}
\end{document}